# Ultraviolet Imaging and Spectroscopy of LINERs

Dan Maoz

*School of Physics & Astronomy and Wise Observatory, Tel-Aviv University, Tel-Aviv 69978, ISRAEL*

**Abstract.** I review the UV properties of LINERs, based mostly on the recent *HST* UV imaging survey of nearby galaxies by Maoz et al. (1995). 25 of the galaxies in the northern subsample host a LINER nucleus, based on the optical spectroscopy of Ho et al. (1996b). Six of these display a prominent compact ($<$ few pc) nuclear UV ($\sim 2300$Å) source in the *HST* images. The remaining 19 LINERs are "UV-dark", with no detectable compact nuclear source. In the six UV-bright objects, the UV flux is correlated with H$\alpha$ flux. When extrapolated beyond the Lyman limit, the UV luminosity is sufficient to produce the observed H$\alpha$ luminosity through photoionization, even without invoking reddening. Some LINERs therefore have a UV continuum source consistent with the expectations from the micro-quasar hypothesis. The 19 UV-dark objects are comparable in H$\alpha$ flux and luminosity to the UV-bright objects, and their darkness is not a detection-limit problem. The UV-dark-LINER host galaxies are on average more edge-on than those of the UV-bright LINERs, but extinction by the host galaxy disk cannot explain the entire effect. There is also no great difference in the Balmer decrements of the two populations, arguing against extinction by foreground dust. The optical line ratios of the two populations are similar as well. I consider several hypotheses explaining why only $\sim 25\%$ of LINERs display a central UV source: Obscuration of the UV source by NLR dust or a molecular torus; the UV source is "turned-off" most of the time; photoionization by an extended population of old stars; or, most LINERs are not photoionized objects. I discuss observational tests that will soon discriminate between these possibilities.

## 1. Introduction

As has been emphasized in this workshop, much of the motivation to understand LINERs is in their commonness; LINER nuclei exist in most early-type spirals (see Ho, Filippenko, & Sargent 1996b and Ho, these proceedings, for the definitive determination of these statistics). If LINERs are indeed "micro-quasars", the AGN phenomenon, albeit in a highly scaled-down version, is a routine element of galactic structure, and can be studied in the nearest galaxies. The same commonness would make

LINERs important for understanding the demise of luminous quasars between $z \sim 2$ and the present, the contribution of low-luminosity AGN to the X-ray background, and the nature of AGNs in general (see Filippenko, these proceedings).

Despite over two decades of debate, there has been no general agreement as to whether or not at least some LINERs are quasars scaled down in luminosity. It is clear now that a LINER-like optical spectrum can be emitted in some non-AGN circumstances, such as cooling flows or ionization by some extended old stellar populations (see beautiful example by Heckman, these proceedings). But are *any* LINERs AGNs, and if so which? It has been long recognized that progress could be made by studying LINERs in additional wavebands, particularly in the UV. One of the clearest signatures of an AGN is a compact source of "nonstellar" continuum emission. Direct detection of this component in LINERs would constitute substantial support for the AGN hypothesis. Such a photoionizing continuum, which in the optical is swamped by light from the old stellar population in the nucleus, would be much easier to detect in the UV. Furthermore, the observation and measurement of UV emission lines can give powerful new diagnostics on the conditions of the line-emitting gas, while absorption features can reveal information on stellar populations that may be emitting the continuum (e.g. Shields 1994). *IUE* has proved of limited use to the UV study of LINERs, because of its large spectrograph aperture (which can admit much diluting starlight) and its low sensitivity (see Reichert 1993, for a review).

The advent of the *Hubble Space Telescope* (*HST*), with its high-angular and spectral resolution and high sensitivity spectroscopic and imaging capabilities, opens many fresh new avenues for deciphering LINERs. The power of UV imaging with *HST* in revealing AGN-like UV sources in galactic nuclei, undiluted by light from the old nuclear stellar population, was demonstrated by Maoz et al. (1995; M95). I outline below the main details of this program and its results.

## 2. The *HST* UV Imaging Snapshot Survey of Nearby Galaxies

M95 carried out a UV imaging survey with the *HST* Faint Object Camera (FOC) of the central regions of 105 large nearby galaxies, selected randomly from a complete sample of 213 galaxies. Full details of the sample selection, observations, and data reduction are given in M95, including some of the results presented here on LINERs. A study of star-forming circumnuclear rings in some of the galaxies is presented in Maoz et al. (1996a). An atlas of the survey is given by Maoz et al. (1996b). The principal purpose of this program is to detect low-luminosity AGNs in the UV, and to study them at small physical scales with *HST*'s high angular resolution. By observing at $\sim 2300$Å, the survey images exclude most of the light from the old bulge population, and isolate AGNs and starforming regions.

The sample of galaxies consists of all galaxies in the UGC and ESO catalogs (Lauberts & Valentijn 1989) with heliocentric velocities less than 2000 km s$^{-1}$ and photographic diameters (as defined in the catalogs) greater than 6'. Exclusion of 57 galaxies for a number of reasons (see M95, Maoz et al. 1996b) left 213 galaxies from which the Space Telescope Science Institute (STScI) staff chose Snapshot targets based only on scheduling convenience. A total of 105 out of the 213 sample galaxies were successfully observed while the program was active.

Data were obtained with the *HST f/96* FOC (Paresce 1990) in its "zoomed" 1024×512-pixel mode with 0.022″×0.044″ pixels, giving a field of view of 22″ × 22″. The F220W filter was used. This is a broad-band filter with an effective wavelength of ∼ 2300 Å and effective bandpass of ∼ 500 Å. The exposure time was 10 minutes per galaxy. The images were processed by STScI's "pipeline" reduction (Baxter et al. 1994). All the M95 data were obtained before the *HST* repair mission at the end of 1993, and therefore are affected by spherically aberrated optics. As a result, the point-spread function (PSF) consists of a sharp core of full width at half maximum (FWHM) ∼ 0.05″ that contains about 15% of the light, with the rest of the light spread in a complex low-level "halo" of several arcsecond radius (Burrows et al. 1991). In the observing mode used, the FOC is limited in its dynamic range to 255 counts (8 bits) per zoomed pixel; additional signal causes the counts to "fold over" and start again from 0. Another problem is that the detected count rate becomes nonlinear, gradually saturating for bright sources (see Baxter et al. 1994). The central pixels of all of the compact bright sources detected in the images are suspect of some nonlinearity, and the brightest of them are clearly saturated. Measurements of brightness and size therefore relied mainly of the halo on the PSF, which has low count rates.

## 3. UV-Bright and UV-Dark LINERs

The FOC images show a variety of morphologies and UV brightnesses in the centers of the galaxies. M95 analyzed the data for nine galaxies whose images show conspicuous bright and compact sources at the galaxy nucleus. Based on the high-resolution ground-based optical spectroscopy described in Filippenko & Sargent (1985) and Ho et al. (1995, 1996b), and some additional spectroscopic observations described in M95, four of the nine galaxies have HII-like nuclear spectra, and five are LINERs. These are among the first clear detections of the long-sought AGN-like continuum source in LINERs. Among the northern-hemisphere galaxies in the sample, 24 are classified as LINERs by Ho et al. (1996b). Some of the galaxies were excluded from the original sample (see above) because they were to be UV-imaged as part of other *HST* programs. I have analyzed existing archival *HST* data for those galaxies. None are LINERs, except M87, which has a bright unresolved nuclear UV source, like the other UV-bright LINERs in the sample. Curiously, although half of the galaxy sample is in the south, all six UV-bright LINERs in the sample are in the north. The final tally is therefore that, out of 25 optically classified LINERs among the northern galaxies of the *HST* sample, six are UV-bright.

Table 1

**UV-Dark LINERs**

| NGC | $f(H\alpha)^1$ | $f_\lambda(2300\text{Å})^2$ | $D$ (Mpc) | $i$ (deg.) | Comments |
|---|---|---|---|---|---|
| 185 | 2.39 | – | 0.7 | 37 | |
| 2768 | 27.88 | – | 23.7 | 67 | |
| 3079 | 15.00 | – | 20.4 | 88 | |
| 3368 | 59.09 | – | 8.1 | 50 | |
| 3627 | – | – | 6.6 | 65 | |
| 3718 | 95.00 | – | 17.0 | 66 | Broad H$\alpha$ comp. |
| 3953 | 13.94 | – | 17.0 | 61 | |
| 4192 | 96.53 | – | 16.8 | 83 | |
| 4216 | 39.36 | – | 16.8 | 89 | |
| 4438 | 667.81 | – | 16.8 | 78 | Extended emission |
| 4636 | 25.14 | – | 17.0 | 44 | Broad H$\alpha$ comp. |
| 4866 | 36.43 | – | 16.0 | 86 | |
| 5005 | – | – | 21.3 | 64 | |
| 5195 | – | – | 9.3 | 46 | |
| 5322 | – | – | 31.6 | 51 | |
| 5566 | 37.73 | – | 26.4 | 71 | |
| 5866 | 10.89 | – | 15.3 | 67 | Broad H$\alpha$ comp. |
| 7331 | 36.64 | – | 14.3 | 68 | |
| 7814 | 3.35 | – | 15.1 | 68 | |

**UV-bright LINERs**

| NGC | $f(H\alpha)^1$ | $f_\lambda(2300\text{Å})^2$ | $D$ (Mpc) | $i$ (deg.) | Comments |
|---|---|---|---|---|---|
| 404 | 106.50 | 1.80 | 2.8 | 0 | |
| 4486 | 44.00 | 0.40 | 16.8 | 47 | M87 |
| 4569 | 441.00 | 10.00 | 12.3 | 64 | |
| 4579 | 129.00 | 1.10 | 20.0 | 36 | Broad H$\alpha$ comp. |
| 4736 | 67.25 | 0.43 | 4.3 | 33 | Double UV source |
| 5055 | 60.00 | 1.00 | 7.2 | 55 | Resolved UV source |

[1] In units of $10^{-15}$ erg s$^{-1}$ cm$^{-2}$. Data compiled from literature and from Ho et al. (1996b)
[2] In units of $10^{-15}$ erg s$^{-1}$ cm$^{-2}$ Å$^{-1}$. For UV non-detections, upper limit is $7 \times 10^{-18}$ erg s$^{-1}$ cm$^{-2}$ Å$^{-1}$.

Table 1 lists the UV-bright and UV-dark LINERs, and some of their properties. Except for NGC 5055, which is marginally resolved (FWHM$\sim 0.2''$), the main UV source in each of the LINERs is unresolved (FWHM$< 0.1''$), giving upper limits of several pc for the size of the UV-continuum emitting region. The result above prompts the question of why only $6/25 \approx 25\%$ of LINERs display a compact central UV point-source. Is the observed UV flux of the right magnitude to produce, through photoionization, the observed emission-line flux? If so, might the expected UV flux of the UV-dark LINERs simply be below the detection limits?

Figure 1 plots the H$\alpha$ flux of the UV-bright (filled circles) and UV-dark (empty circles) LINERs vs. their observed UV flux (or upper limit). The solid line shows the H$\alpha$ flux produced by a $\nu^{-1}$ extrapolation beyond the Lyman limit of the observed UV flux, assuming Case B recombination and a covering factor of 1. All six UV-bright LINERs are on or to the right of the line, implying they have more than the minimum UV flux required to produce H$\alpha$ at the observed strength. Any correction for reddening will reinforce this conclusion, as it will move the points more to the right than upwards. This sufficiency test and the correlation of the H$\alpha$ and UV fluxes is consistent with the hypothesis that the UV-bright LINERs are photoionized. Except for one or two objects, the UV-dark LINER upper limits on the UV flux are far to the left of the solid line, so their darkness is not a detection-limit problem.

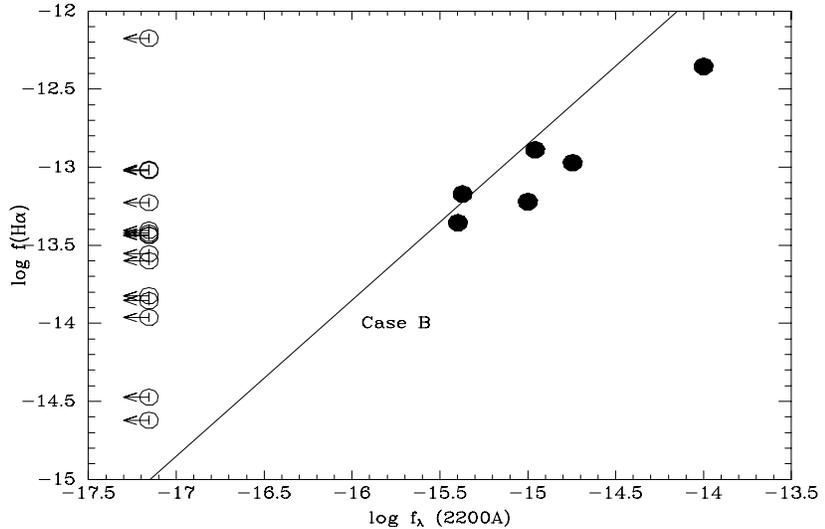

Figure 1. H$\alpha$ flux of the UV-bright (filled circles) and UV-dark (empty circles) LINERs vs. their observed UV flux (or upper limit). The solid line shows the H$\alpha$ flux produced by a $\nu^{-1}$ extrapolation beyond the Lyman limit of the observed UV flux, assuming Case B recombination and a covering factor of 1. All six UV-bright LINERs are on or to the right of the line, implying they have more than the minimum UV flux required to produce H$\alpha$ at the observed strength.

Figure 2 shows the same data, but with luminosities (somewhat uncertain because of the uncertain distances to such nearby galaxies) instead of observed fluxes. Here we also see that the UV-dark and UV-bright LINERs are not significantly different in their ranges of H$\alpha$ luminosity. An obvious way to hide the UV sources in the UV-dark LINERs would be through dust obscuration. The dust could conceivably be in the disk of the host galaxy, in patches or filaments surrounding the nucleus,[1] mixed in with the narrow-line-region (NLR) gas, or in a pc-scale molecular torus that hides only the continuum source (but does not affect the lines). As a way of testing the first (galaxy disk) possibility, I plot in Figure 3 the distribution of host galaxy inclination for the two populations. On the one hand, the UV-dark LINER galaxies are on average more edge-on than the UV-bright ones (see also contribution by Barth, these proceedings). On the other hand, there is a large overlap in the two distributions, so obscuration by the galaxy disk is unlikely to explain the whole effect. As a test of *any* foreground reddening, I show in Figure 4 the distributions of Balmer decrement (as measured by Ho et al. 1996b). Again, some of the UV-dark objects are highly reddened, but there is a general overlap between the two populations. The two galaxies with the largest Balmer decrement in Fig. 4 are two of those with the largest inclination in Fig. 3. The conclusion is that a few of the LINERs are highly reddened by foreground dust, which could hide a UV source if it were there, but that this mechanism cannot account for most of the UV-darkness effect. In §4 I outline tests for dust at the other locations noted above.

---

[1] Van Dokkum & Franx (1995) show, based on *HST* archival images, that dust with a variety of morphologies exists in the centers of most galaxies. See also contribution by Barth (these

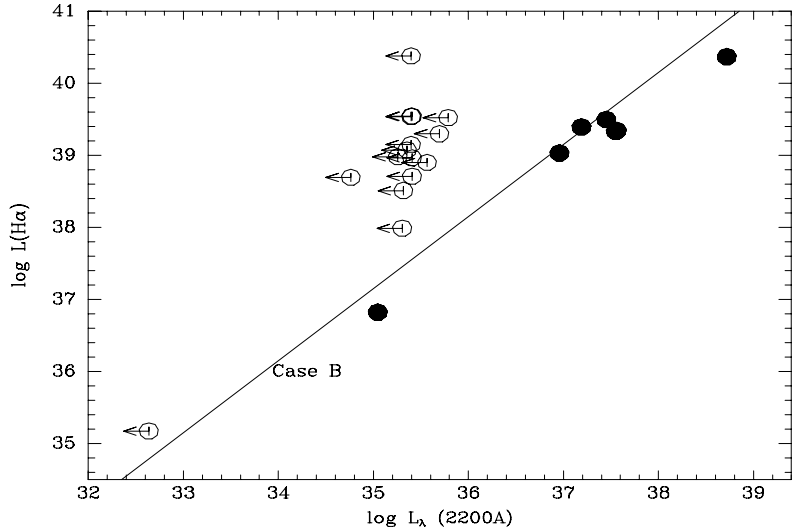

Figure 2. As in Figure 1, but with luminosities instead of observed fluxes. $L_\lambda$ in units of erg s$^{-1}$ Å$^{-1}$.

Could there be intrinsic differences between the two populations that are reflected in other properties? A useful tool in LINER classification and analysis has been diagnostic diagrams that compare various emission-line ratios (e.g. Veilleux & Osterbrock 1987). Figure 5 shows the UV-dark

---

proceedings) with spectacular *HST* images showing dust filaments criss-crossing some galactic bulges.

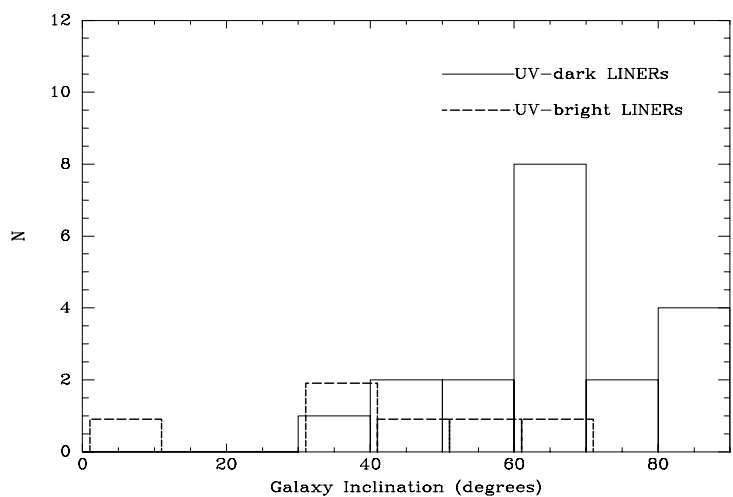

Figure 3. The distribution of host galaxy inclinations for the UV-bright and UV-dark LINERs

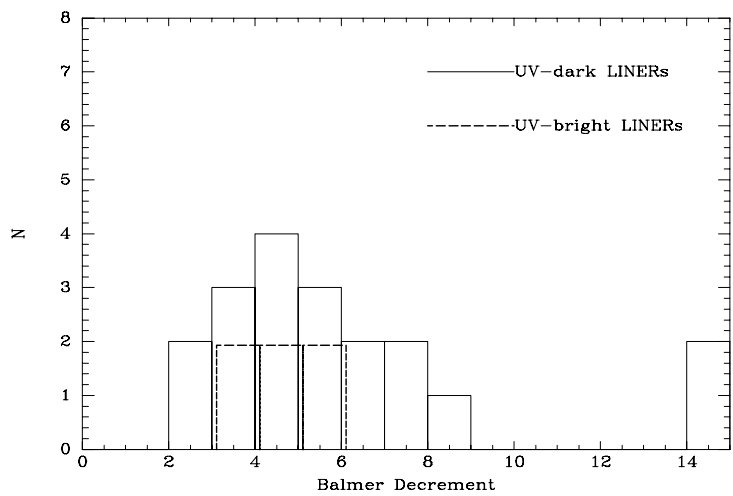

Figure 4. The distribution of Balmer decrements for the UV-bright and UV-dark LINERs

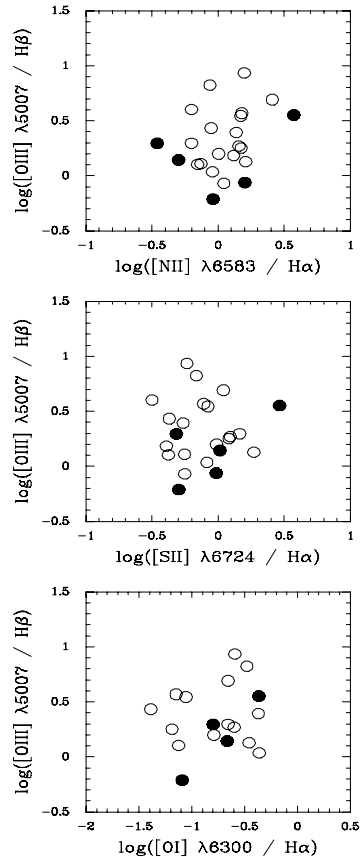

Figure 5. Diagnostic line-ratio diagrams for LINERs common to the *HST* sample and the Filippenko & Sargent (1985) sample. Filled circles denote the LINERs that contain detected nuclear UV sources, empty circles show the other, UV-dark LINERs. No obvious segregation exists between the UV-bright and UV-dark LINERs, suggesting that they may be one class of objects.

and UV-bright LINERs on three such diagrams, based on the Ho et al. (1996b) optical measurements. There is no clear segregation between the open circles (UV-dark) and filled circles (UV-bright). Thus, whatever the physical cause for the difference between UV-bright and UV-dark LINERs, it does not visibly affect their optical line ratios.

## 4. Possible Explanations and Future Tests

As shown above, the data presently do not support the idea that most UV-dark LINERs are obscured by foreground dust in the host galaxy disk or in dust filaments and patches crossing the nucleus. Dust could be mixed in with the NLR gas (see contribution by Netzer, these proceedings) without significantly raising the Balmer decrement. The observed fraction of UV-dark objects would result if the NLR covering factor were about 75%. A prediction of this scenario is that the emission-line ratios at additional wavelengths (IR, UV) would show the mild reddening effects from this kind of dust-gas mixture, but would be the same for the two populations. The UV-dark LINERs would show larger absorbing columns in X-ray observations.

Another possible location for dust that is hiding the UV is in a torus of molecular gas surrounding the central UV source, with the torus orientation determining the UV properties (e.g. Antonucci 1993). UV-bright/dark LINERs would then be the analogs of Seyferts 1 and 2. The analogy cannot be perfect, because there is no correspondence between the detection of broad H$\alpha$ wings (i.e. a low-luminosity broad-line region [BLR]) and UV-brightness; Among the UV-bright objects only NGC 4579 has faint broad H$\alpha$ wings, and three of the UV-dark ones do have them (Filippenko & Sargent 1985, Ho et al. 1996b). Perhaps BLRs sometimes do not exist, and when they do they can be revealed even though the continuum is blocked. In the torus scenario the emission-line spectrum of both types of LINERs would again be the same at IR and UV bands, while X-ray spectra would show the large absorptions expected from the torus in the UV-dark objects. If a collimating structure like a torus exists, narrow-band emission-line imaging could reveal elongated or conical structures, as are often observed in Seyfert-type AGNs (e.g. Pogge 1989a; Wilson & Tsvetanov 1994) . As shown by Pogge (1989b), ground-based searches for such structures in LINERs have proved inconclusive, with hints of structure at the seeing limit. Figure 6 shows examples of H$\alpha$ images obtained at Lick Observatory for four of the LINERs discussed here. Imaging of LINERs with *HST* could reveal such "ionization cones" at sub-arcsecond scales, and could test if the UV-bright LINERs appear more face-on. An alignment with radio structures mapped with the VLA would provide further support for something akin to the torus model, and would constitute additional evidence linking LINERs to the AGN class.

An alternative to the obscuration explanation is the duty-cycle hypothesis proposed by Eracleous et al. (1995). In this scenario, the emission-lines in LINERs are fluorescing in response to a photoionizing continuum that is in its "on" state only 25% of the time. Such a duty cycle could arise from episodes of tidal disruption and accretion of passing stars by a central massive black hole. Most of the emission-lines would remain largely unchanged in strength during the century-long intervals between accretion events, explaining the lack of spectral difference between UV-bright and UV-dark LINERs. One emission line that would decay quickly though (on time-scales of a few years) would be [OIII]$\lambda$5007. A prediction of this model is that high angular-resolution narrow-band [OIII] imaging of UV-dark LINERs will reveal a "gap" in [OIII] emission closest to the nucleus, corresponding to the region of the NLR transversed by the light-travel-time front, behind which the [OIII] emission has decayed. Accurate (few %) spectrophotometry of the [OIII] flux from UV-dark LINERs should also reveal a decreasing flux on timescales of 5-10 years. Finally, like the obscuration hypothesis, the duty-cycle model also predicts a correspondence between UV-brightness and broad H$\alpha$ wings, since excitation of BLR gas will cease as soon as the ionizing continuum is off. The existence of UV-dark LINERs with broad H$\alpha$ contradicts this prediction.

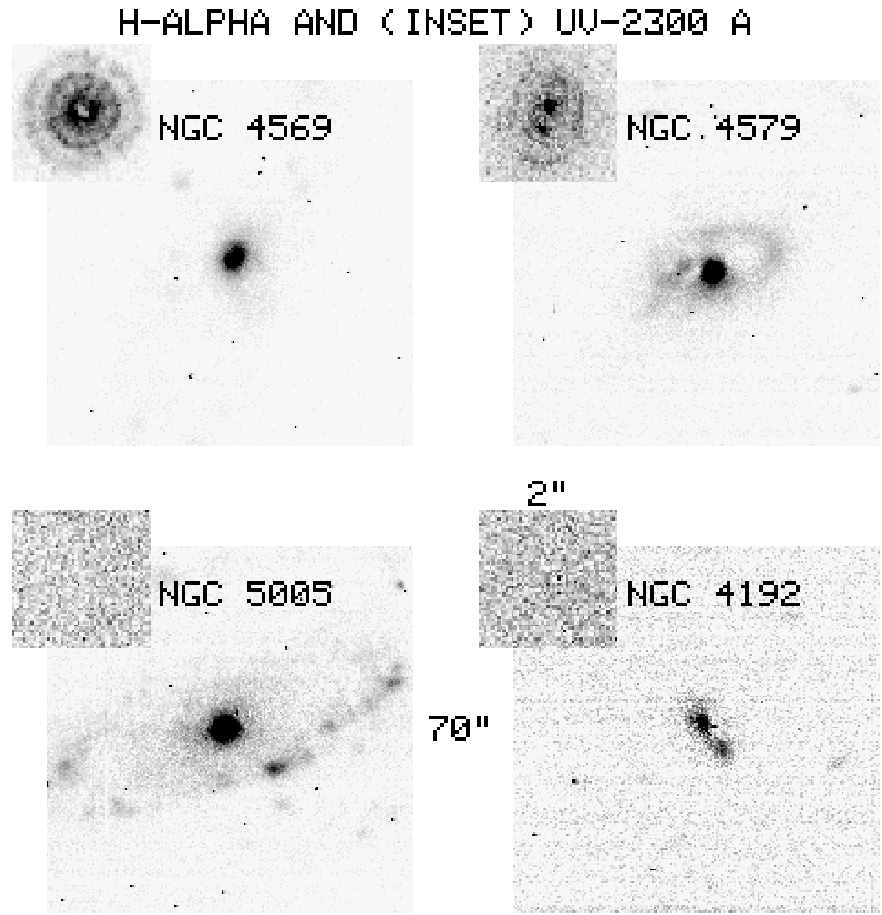

Figure 6. Ground-based H$\alpha$+[NII] emission-line images, with the HST/FOC UV (2300Å) images from Maoz et al. (1995) inset. The top two panels show the UV-bright LINERs NGC 4569 and NGC 4579, while the bottom two show the UV-dark LINERs NGC 5005 and NGC 4192. Angular scales are indicated in the margin on the bottom two panels.

Moving in the direction of explanations suggestions that UV-bright and UV-dark LINERs are *not* intrinsically the same objects, Heckman (these proceedings) has demonstrated that the integrated light from a large region around the nucleus in M31 has a LINER spectrum, although the nucleus itself does not. Presumably, photoionization of interstellar gas by diffuse UV radiation from evolved stars is doing the job (e.g. Binette 1995). The same thing may be happening in some of the UV-dark LINERs; because they are more distant than M31, the same large area would fit into the ground-based spectral aperture and produce a LINER spectrum, but the UV flux would be diffuse enough to be undetected in the FOC images. Furthermore, this UV emission from evolved stars (the "UV-upturn") may set in at wavelengths shorter than covered by the F220W band. Narrow-band H$\alpha$ imaging with *HST* of UV-dark LINERs would easily show whether or not the emission is diffuse and extended, rather than NLR-like, and so test this explanation for the UV-dark LINERs.

Finally, the absence of a UV source in most LINERs may indicate a true, physical absence, and the presence of a non-photoionizing excitation mechanism. The favorite contender, has of course always been shock excitation (e.g. Heckman 1980), and it still has its advocates (e.g. Dopita & Sutherland 1995; Dopita, these proceedings). Shock and photoionization models will soon have a much better confrontation with observations with the availability of IR spectroscopy from the *ISO* satellite, and UV spectroscopy by *HST*. Here too, comparison of high-resolution radio maps to high-resolution narrow-band images can reveal signatures, such as bow-shock structures.

## 5. UV Spectroscopy

As is evident from the preceeding discussion, UV spectroscopy holds a key not only to the UV-bright/dark question but to the nature of LINERs in general. Keel (1995) has used *IUE* spectra of NGC 4569 (the brightest UV-bright LINER in the *HST* survey), in combination with the M95 FOC UV results and ground-based optical spectroscopy to argue that this LINER must be dominated by a non-stellar component. But it has been clear for a long time that UV spectroscopy with *HST* will give the true push to exploiting the information from this spectral range. *HST* FOS spectra of most of the UV-bright LINERs have been recently, or will soon be, obtained.

In the meantime, the *HST* UV spectrum of only one LINER, M81, has been shown (Ho, Filippenko, & Sargent 1996a). They show that, in terms of its featureless-continuum and the properties of its narrow and broad emission lines, the M81 nucleus is consistent with being a true quasar-like AGN (with some important differences in overall spectral energy distribution.) An *off*-nuclear *HST* UV spectrum of M87, a UV-bright LINER, was shown in this workshop by Mike Dopita. It shows clear signatures of shock excitation, as would be expected from the jet-ISM interaction in this object. The prospect of seeing on-nuclear UV spectra of more LINERs in the near future is exciting.

## 6. Conclusions

The existing UV imaging and spectroscopy data strongly suggest that at least about 1/4 LINERs are photoionized by a compact source of UV emission. The spectroscopic data existing for one or two of these UV-bright objects also point to an AGN-like nature for the central source, rather than a hot young and compact star cluster. It will be interesting to see if this conclusion holds up after UV-spectra of additional UV-bright LINERs are seen.

The nature of the remaining 3/4 of LINERs is an open question. I have outlined a series of tests that can determine whether they are also photoionized by an AGN (that is obscured or turned off), or photoionized by old stars, or excited by some other mechanism, presumably shocks. Many of these tests will soon be carried out, taking advantage of the angular resolution of *HST*, and the new spectral ranges in which LINERs are being explored for the first time thanks to *ISO*, *HST*, and *ASCA*. The picture will hopefully be clearer soon. It is likely that a combination of factors are behind the UV-darkness of most LINERs; for example, the emission-lines in NGC 4438, a UV-dark LINER in Table 1, are extended (Filippenko & Ho, private communication), and probably arise in a cooling flow. NGC 3079 and NGC 4192, also in Table 1, are reddened enough that a UV source would probably be invisible even if it were there.

The upcoming observations will enable us to cull such exceptions from the UV-bright and UV-dark lists, and then get a better idea of what the true fractions are, and whether they can be unified within some model. But even if only the 25% or so of LINERs that are UV-bright are indeed AGNs (as the first spectra seem to indicate), it would mean that micro-quasars are very common in nearby galaxies.